\documentclass[%
 reprint,
 amsmath,amssymb,
 aps,
floatfix,
]{revtex4-2}

\usepackage{graphicx}
\usepackage{dcolumn}
\usepackage{bm}

\usepackage[dvipsnames]{xcolor}

\usepackage{accents}

\usepackage{ulem}

\begin{document}

\preprint{APS/123-QED}

\title{Park visitation and walkshed demographics in the United States\\}

\author{Kelsey Linnell}
\altaffiliation{%
Vermont Complex Systems Center,
 University of Vermont 
}%
\author{Mikaela Fudolig}%
\altaffiliation{%
Vermont Complex Systems Center,
 University of Vermont 
}%
\author{Laura Bloomfield}
\altaffiliation{Gund Institute for Environment, University of Vermont}
\author{Thomas McAndrew}
\altaffiliation{College of Health, Lehigh University, Bethlehem, PA, United States}
\author{Taylor H. Ricketts}%
\altaffiliation{Gund Institute for Environment, University of Vermont}
\author{Jarlath P. M.
O’Neil-Dunne}%
\altaffiliation{Gund Institute for Environment, University of Vermont}
\author{Peter Sheridan Dodds}%
\altaffiliation{%
Vermont Complex Systems Center,
 University of Vermont 
}%
 \author{Christopher M. Danforth}%
\altaffiliation{%
Vermont Complex Systems Center,
 University of Vermont 
}%

\date{\today}

\begin{abstract}

A large and growing body of research demonstrates the value of local parks to mental and physical well-being. 
Recently, researchers have begun using passive digital data sources to investigate equity in usage; exactly who is benefiting from parks?
Early studies suggest that park visitation differs according to demographic features, and that the demographic composition of a park's surrounding neighborhood may be related to the utilization a park receives. 
Employing a data set of park visitations generated by observations of roughly 50 million mobile devices in the US in 2019, we assess the ability of the demographic composition of a park's walkshed to predict its yearly visitation. 
Predictive models are constructed using Support Vector Regression, LASSO, Elastic Net, and Random Forests. 
Surprisingly, our results suggest that the demographic composition of a park's walkshed demonstrates little to no utility for predicting visitation. 

\end{abstract}

\maketitle


\section{\label{sec:intro} Introduction}

The positive impact of park access, proximity, and use on well-being has been well established.
Populations with park access have been found to be more active, have lower rates of obesity, and overall better cardiovascular health \cite{reuben_association_2020, lee_proximity_2019, macfarlane_modeling_2021, stark_impact_2014}.
As a form of nature exposure, park access also benefits cognition and mental well-being and is associated with lower stress and reduced rates of depression \cite{orsega-smith_interaction_2004, bojorquez_urban_2018, roe_green_2013, bratman_impacts_2012}. Independent of time spent in the park, residential proximity is positively correlated with improved mental health \cite{sturm_proximity_2014}.

Inequity in park resources can arise as a function of accessibility, or in qualitative differences in parks \cite{dai_racialethnic_2011, rigolon_complex_2016, rigolon_inequities_2018, wolch_parks_2005, zhou_social_2013, rigolon_parks_2017, wolch_urban_2014}.
Studies have found that even when spatial access is equitable, income and race are linked to park quality.
Parks associated with whiter and more affluent parks have more acreage, more tree canopy, and different amenities \textemdash including more playgrounds  \cite{rigolon_parks_2017, rigolon_inequities_2018, zhou_social_2013, vaughan_exploring_2013, choi_xs_2020, locke_residential_2021}. 

Differences in park quality are significant because they may drive differences in visitation, which is the assumed mechanism by which parks offer health benefits.
For example, while studies find correlations between residential proximity and lower rates of obesity, the underlying mechanism assumed to create this correlation is often that proximity to a park increases the likelihood of exercise in the park \cite{cohen_contribution_2007,huang_neighborhood_2020,knapp_relationships_2019, alajajian_lexicocalorimeter_2017}.
Similarly, studies on the impact of green space on mental health indicate that benefits are either accessed or increased by visitation to the space \cite{orsega-smith_interaction_2004, yuen_changes_2020, schwartz_visitors_2019}.
Park-based physical activity, for example, can mediate the relationship between park proximity and mental health benefits \cite{orstad_park_2020}.
Thus, realized usage, or visitation, is an important variable when studying equity and the heath impact of parks~\cite{saxon_empirical_2021}.
Measuring park usage is particularly useful in equity studies because it can be used to quantify the effect of non-geographic barriers. 

Non-geographic barriers to access---such as time constraints and safety ---create inequity in the benefits that communities receive from parks.
A much studied example of such a barrier is the perceived safety of parks \cite{groshong_attitudes_2020, groshong_exploring_2017, lapham_how_2016, cohen_paradox_2016, orstad_park_2020, kamel_disparities_2014}.
Scholars have found that parks in higher income areas had fewer safety concerns than parks in low income areas and that perceptions of park safety are strongly tied to the odds of visiting a park~\cite{kamel_disparities_2014,lapham_how_2016}. 
The impact of perceived safety as a barrier is visible in a 2020 study by Orstad, where mental health benefits were only associated with living in proximity to a park for residents who did not have concerns about park crime \cite{orstad_park_2020}. 

Classifying non-geographic barriers, and measuring their impact on communities remains an active area of research. Investigations with regards to these barriers requires measuring realized usage.
Historically, park visits have been quantified using either surveys or by in-person observation of visitors \cite{ho_gender_2005,stark_impact_2014,sturm_proximity_2014,wang_comparison_2015,rivera_important_2021,lapham_how_2016,cohen_paradox_2016,cohen_contribution_2007,huang_neighborhood_2020}.
These methods are limited both geographically and temporally, and can be expensive to implement.

Recently, these temporal and geographic limitations have been managed by using digital data sources such as social media and GPS data from mobile devices \cite{tenkanen_instagram_2017, hamstead_geolocated_2018, donahue_using_2018,schwartz_visitors_2019, schwartz_gauging_2022}.
Digital data sources offer the ability to measure park usage at all times of the day, for prolonged periods of time, over a large geographic scale.
Using over 2 million observations of smartphone use, texting, calling, and environmental exposures for 700 young adults, Minor et al. saw calling and texting actually increase during short recreational greenspace visits \cite{doi:10.1177/00139165231167165}.
Longer exposures led to smartphone use decline however, suggesting that nature exposure may support digital impulse inhibition. Importantly, participants with elevated baseline screen-time significantly reduced device use in nature.

By using Twitter and Flickr data, Hamstead et al. \cite{hamstead_geolocated_2018} were able to observe differences in visitation to all of the parks in New York City; their findings suggest visitation varies based on not only characteristics of the park, but on the demographic composition of the neighborhood surrounding the park. If park visitation varies with the demographic composition of its neighborhood for the US in general, it may be possible to predict park visitation using demographic data from its neighborhood.

Predicting which parks are being underutilized could inform decisions about which parks need more investment, better infrastructure, new programming, or could benefit from investigative studies into the barriers preventing full utilization. To explore whether park visitation can be predicted from demographic features, we use a novel dataset of daily park visitation counts obtained through observations of approximately 50 million mobile devices for 2,506 parks throughout the contiguous United States. 
In particular, we focus on the population residing within a ten minute walk of the park.
This area is referred to in the literature as a ``walkshed" and represents the residential area for which a park is considered “accessible” \cite{hamstead_geolocated_2018, sandalack_neighbourhood_2013, miyake_not_2010} 

Historically, residents of the walkshed have been considered the primary users of the park, and walksheds have been conceptualized of as a fixed radius buffer around the park.
Here we establish a park's walkshed using the convex hull of a pedestrian walking network within ten minutes walk of a park boundary. Using this method, the walkshed is limited to areas that have a walkable route to the park, and parks which cannot be reached by foot are excluded. 
Employing census data, we attribute demographic characteristics to the residents of a park's walkshed, and evaluate the ability of these characteristics to predict the visitations received by the park itself. Echoing the literature, we explore dimensions of race, income, educational attainment, gender, and age on park visitation in the United States.

\section{\label{sec:Data} Data}

\subsection{\label{sec:park_data}Park Visitation Data}

Our data set consists of daily records of the number of unique visitors to 7,997 non-commercial parks in the contiguous United States for the year of 2019.
Non-commercial parks include city, municipal, and neighborhood parks. 
Because we wish to focus on parks which are frequented by those living near them, we excluded national and state parks which often attract tourists and infrequent visitors from a wide geographic area.  

The number of daily visitors to each park was estimated using the number of unique mobile devices that reported GPS data from within the park bounds on that day to the company UberMedia (since acquired by Near)~\cite{whitepaper}. 

UberMedia---a data vendor specializing in GPS data acquired from cellphones--collects GPS coordinates with timestamps from billions of devices.
GPS coordinates are collected when a person uses one of over 400 location collecting apps or when they are exposed to advertising through real-time bidding.

In 2019 the average number of unique devices that UberMedia observed each day varied between 44 and 60 million, representing approximately 10\% of the adult population in the United States~\cite{whitepaper}.

\begin{figure}
\includegraphics[width=.5\textwidth]{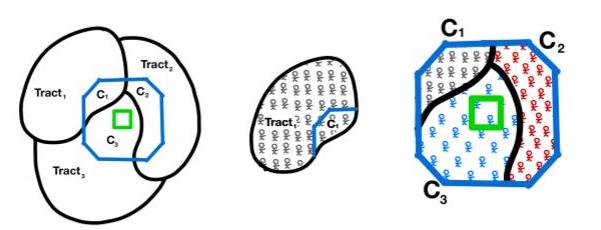}
\caption{\label{fig:cartoon} \textbf{A conceptual model of the walkshed.} In the left most image the park is represented by a green square, and the walkshed is represented by the blue polygon surrounding the green square. The three black bordered shapes labeled ``Tract" demonstrate how a walkshed could intersect multiple Census Tracts. The intersections of the walkshed with the census tracts are labeled $C_i$, where $C$ indicates the region is a component of the walkshed, and $i$ refers to the census tract that the particular component lies within. The middle image demonstrates the assumed uniform spatial distribution of a homogenous population within the tract, where the people associated with the component have the same features as the tract, and the population is proportional to the area of the census tract covered by the component. The final image is of the walkshed, with each of its components populated with respect to the census tract in which they lie. The total walkshed population is considered to be the aggregation of the populations of each component. }
\end{figure}

\subsection{\label{sec:census_data}Census and Geographic Data}

Demographic data were retrieved from the 2019 American Community Survey 5 year data using the Census API.
Census Tracts were chosen as the geographic unit because they are designed to be relatively homogeneous units with respect to the demographic features of their residents.
We collected the following demographic data: race, ethnicity, age, median income, and educational attainment of residents over 25.

Demographic data were linked to park visitation data generated by Ubermedia based on the intersection between census tracts boundaries and areas within walking distance of the park~(which we term a \textit{walkshed}).
The process of defining walksheds and linking demographic data to parks is described in section \ref{sec:walkshed}.

\section{\label{sec:Methods} Methods}

Exploring the predictive value of the demographic features of the surrounding neighborhood for a park's visitation required joining the data sets mentioned in \ref{sec:Data}, confirming the suitability of the resulting data for the study, and evaluating the predictive ability of several models.

Each of these steps are detailed in the subsections below.
Section ~\ref{sec:walkshed} explains how demographic data from the US Census were joined to park visitation data using a geographic unit called a walkshed.
This section also discusses how population data were aggregated for the walkshed.
The inclusion criteria chosen for the study is delineated in section ~\ref{sec:inclusion_criteria}. The resulting data set is described in section ~\ref{sec:study_set}. Compared to the US population, the walksheds included in our data have a large distribution of demographic features.
Finally, section ~\ref{sec:analysis} explains the methodology by which models were chosen, fit, and evaluated.

\begin{figure*}
    \centering
    \includegraphics[width=\textwidth]{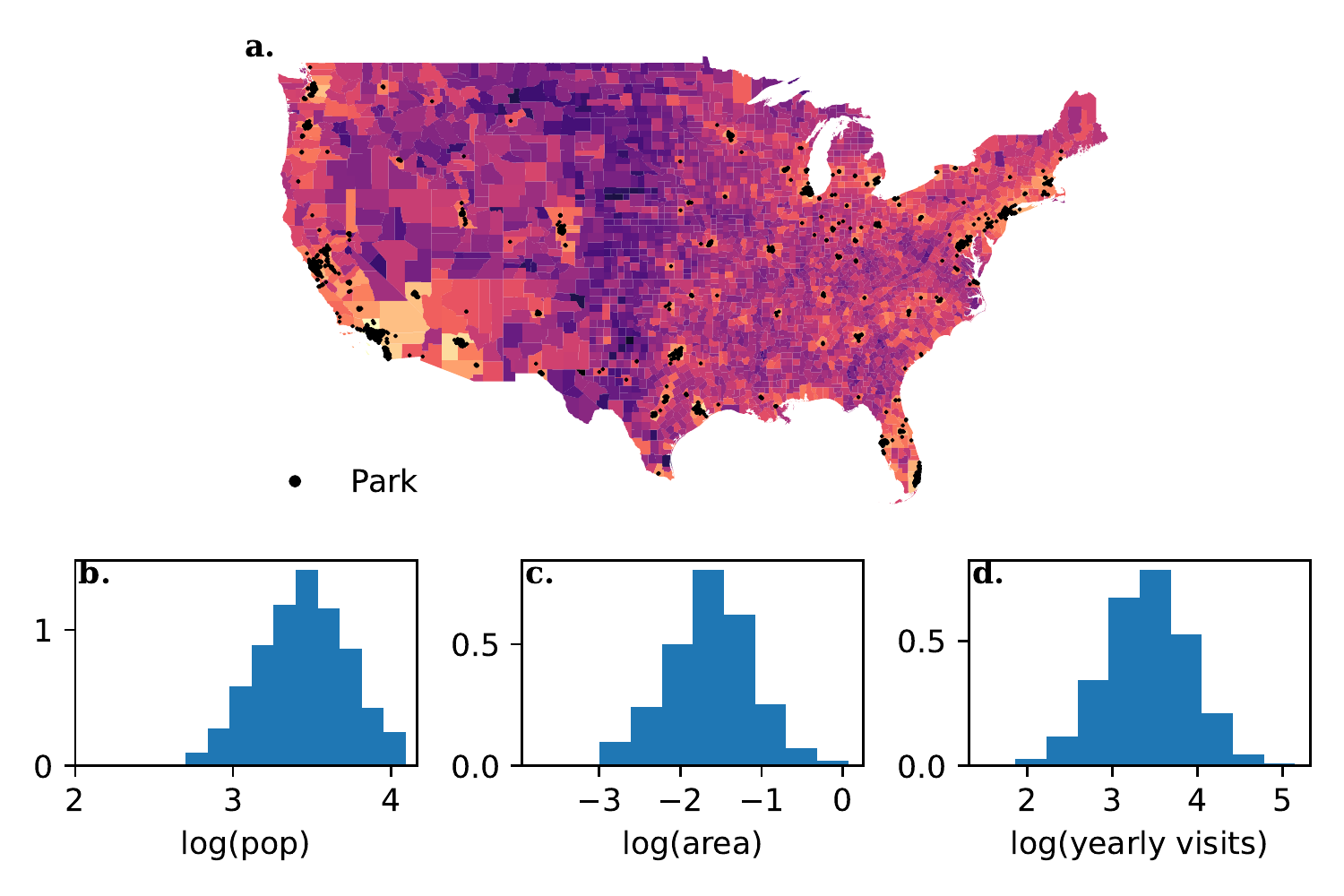}
    \caption{\label{fig:map}\textbf{Heat map of the population of the contiguous United States overlaid with the locations of the parks used in the study, with each park demarcated by a black point, and distributions of park sizes and popularity below}.
    Histograms display the log normal distributions of parks in the dataset across walkshed population, park area, and yearly visits. The median park has 2785 people in its walkshed, covers 5.68 acres, and received 2528 visits in 2019. }

\end{figure*}

\subsection{\label{sec:walkshed} Walkshed Construction}

For each of the parks in our dataset we constructed a walkshed and calculated an estimated walkshed population. 
The walkshed was defined to be the convex hull of the graph of the walking network obtained from OpenStreetMaps that represented a ten minute walk to the boundary of the park.
This walkshed was then considered to be composed of the disjoint components lying in unique Census Tracts (see \ref{sec:census_data}). Census Tracts are treated as homogenous populations uniformly distributed over a geographic area.
Walksheds for each park in our study were created through a series of steps (Figure \ref{fig:cartoon}).
The estimated number of people in a walkshed associated with park $p$,
$P^{p}_{\text{walkshed}}$ was computed as 

\begin{equation}
    P^{p}_{walkshed} = \sum_{i=0}^{n} P_{tract_i} \frac{A_{i}}{A_{tract}} %
\end{equation}
\label{eq:one}

where $P_{tract_i}$ is the estimated number of residents in the $i^{th}$ census tract intersecting with the walkshed, ${A_i}$ is the area of that intersection, and $n$ is the total number of census tracts intersecting with the walkshed.

In addition to the number of people in the walkshed, we estimated the income, gender, age, educational attainment, ethnicity, and racial composition of the walkshed population. The income of a person within the walkshed was calculated as the population-weighted average of the median incomes of the tracts intersecting that walkshed:

\begin{equation}
I_{walkshed} =\frac{1}{P_{walkshed}} \sum_{i=0}^{n} P_i  I_{i}%
\end{equation}
\label{eq:three}

where $I_{walkshed}$ is our estimate of the  median income of a person in the walkshed and $I_{i}$ is the median income of the census tract containing component $i$.

The average age of the residents was calculated similarly. The proportion of the walkshed belonging to a racial group, ethnicity, age range, sex, or having reached a given educational attainment was calculated as:

\begin{equation}
 P_{walkshed,j} =\frac{1}{P_{walkshed}} \sum_{i=0}^{n} P_{i,j} %
\end{equation}
\label{eq:four}.

Where $i$ refers to the component, and $j$ refers to the classification of interest, for example $j \in \{ race_{1}, race_{2}, \cdots race_{J}  \} $.

\subsection{\label{sec:inclusion_criteria} Inclusion Criteria}

A park was included in analysis if (i) the walkshed had an area greater than zero (i.e. the park could be accessed on foot) and (ii) demographic data were available from the US census for each census tract intersecting with the walkshed.
Parks were omitted from the study if their walkshed included a component contained in a tract for which the US Census reported no data. 

Because we assumed people in a census tract were distributed uniformly, we excluded a walkshed if it contained a census tract with less than one person per quarter-acre.

Parks with walksheds containing fewer than 500, or more than 12,500 people were also excluded. Parks with high population walksheds were considered large enough to attract tourists. Parks with fewer than 500 residents were considered too rural, with potentially inconsistent visitation. 
These parks accounted for 154 of the parks in the dataset. 
Of the initial 7,997 parks, 2,506 met our inclusion criteria. The primary reason for exclusion was insufficient population density.

\subsection{\label{sec:study_set} Study Set Description}

Our data set contained 2506 parks, which we observed to exhibit log-normal distributions in walkshed population, park area, and yearly visits (Figure ~\ref{fig:park_info}).
The average park in our data set was 0.023 $\textrm{km}^2$ (5.68 acres), contained 2,785 people in its walkshed, and received 2,528 visits each year. 

\begin{figure*}
\includegraphics[width=.9\textwidth]{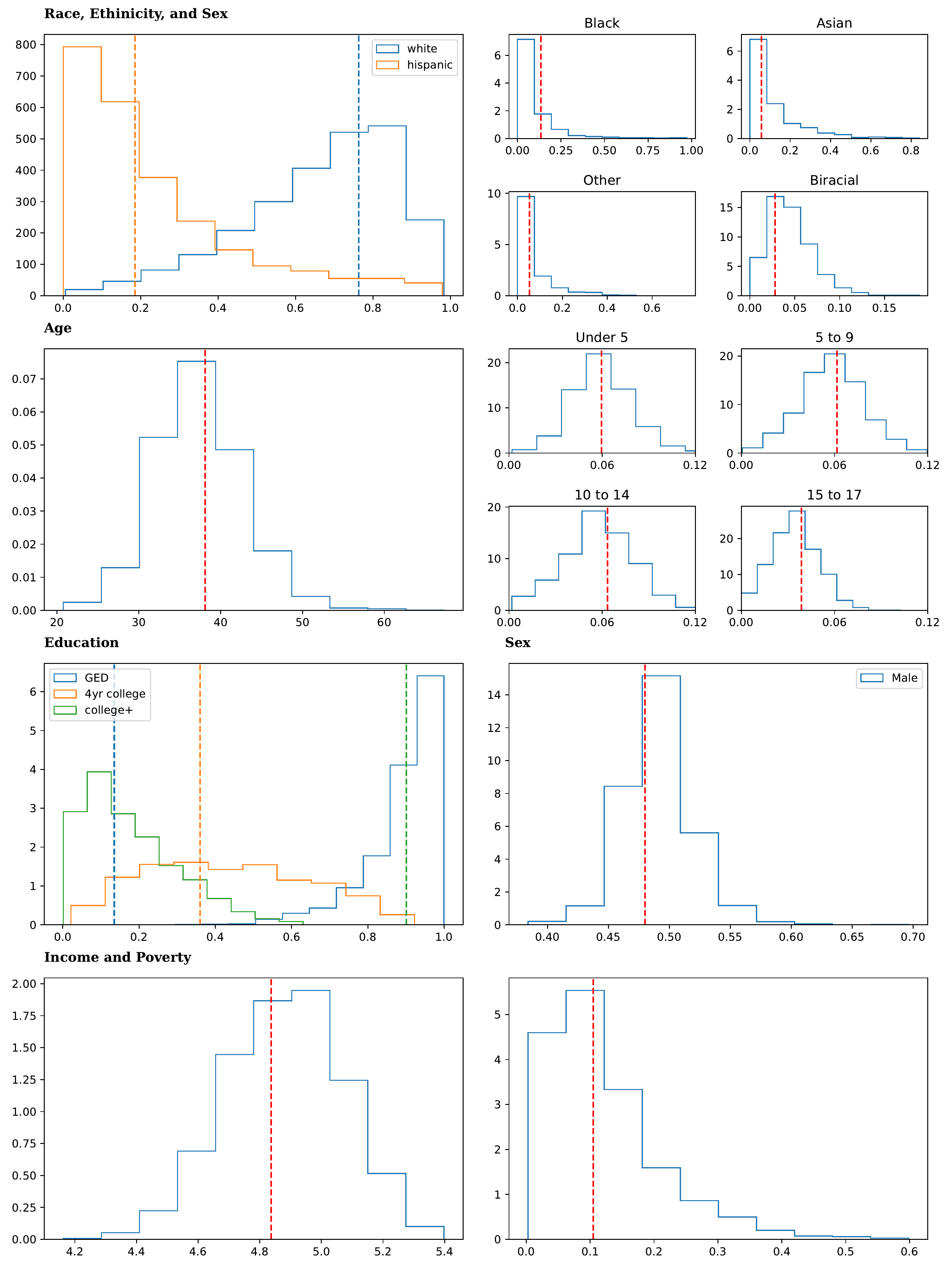}
\caption{\label{fig:demographic_mega} \textbf{Distributions of the parks in the study set across race, ethnicity, sex, educational attainment, age and measures of wealth.} The first row displays the distribution of parks in the study set across fraction of walkshed population in four racial categories (Black, Asian, Multiracial, and White), and fraction identifying as Hispanic. The second row indicates the distribution of parks in the study set across average age -defined as the population weighted average of the medians for each component, and the fraction of walkshed residents falling into each of four child age ranges (under 5, 6-10, 11-14, and 15-17).The third row displays the distribution of parks in the study set across fraction of walkshed population over 25 having earned at least a high school diploma (or equivalent) (blue), at least a Bachelor's degree (orange), and with more than a Bachelor's degree (green), as well as the fraction that is male (right). The final row displays the distribution of walkshed populations across the log of the average household income, and the fraction living below federal poverty level. Dashed vertical lines in each of the plots indicate the corresponding value for the general population of the United States in 2019.}
\end{figure*}

The income associated with walksheds was also log-normally distributed.
The average household income associated with a park’s walkshed was \$76,155 per year, which is \$7,542 more than the US national median income.
Approximately 54\% of communities associated with a walkshed had a lower proportion of individuals living below the Federal Poverty Level than the United States as a whole.

Figure ~\ref{fig:demographic_mega} presents a summary of the distribution of walksheds with regards to race, ethnicity, educational attainment, and age. 55\%  and 46\% of the parks in our dataset had a greater proportions of Asians and Hispanics than in the general population, while only 22\% and 37\%  of parks had greater proportions of Black and White people than the general population.The average proportion of females within a walkshed was
51.0\% compared to 49.8\%, the estimated proportion
of females in the US.

Comparing the educational attainment of walkshed populations to that of the general population, we find that approximately 60\% of the walkshed populations had more highschool graduates, college graduates, and persons with advanced degrees than the general population. Overall, walkshed populations tended to have higher educational attainment than the general population. On average the parks in our data set had a similar median age, and similar proportions of children in each age group, as the general population. 

\subsection{\label{sec:analysis} Analytic Methodology}

In our data set, the area of the park in sqkm and walkshed population are both positively correlated with yearly visitation on a log-log scale (Pearson correlation of 0.61 and 0.47 respectively, Spearman correlation of 0.59 and 0.47).
Both factors effect the spatial accessibility of the park, but do not necessarily reflect the users of the park, or the demographic environment of the park.

\begin{figure*}
    \centering
    \includegraphics[width=\textwidth]{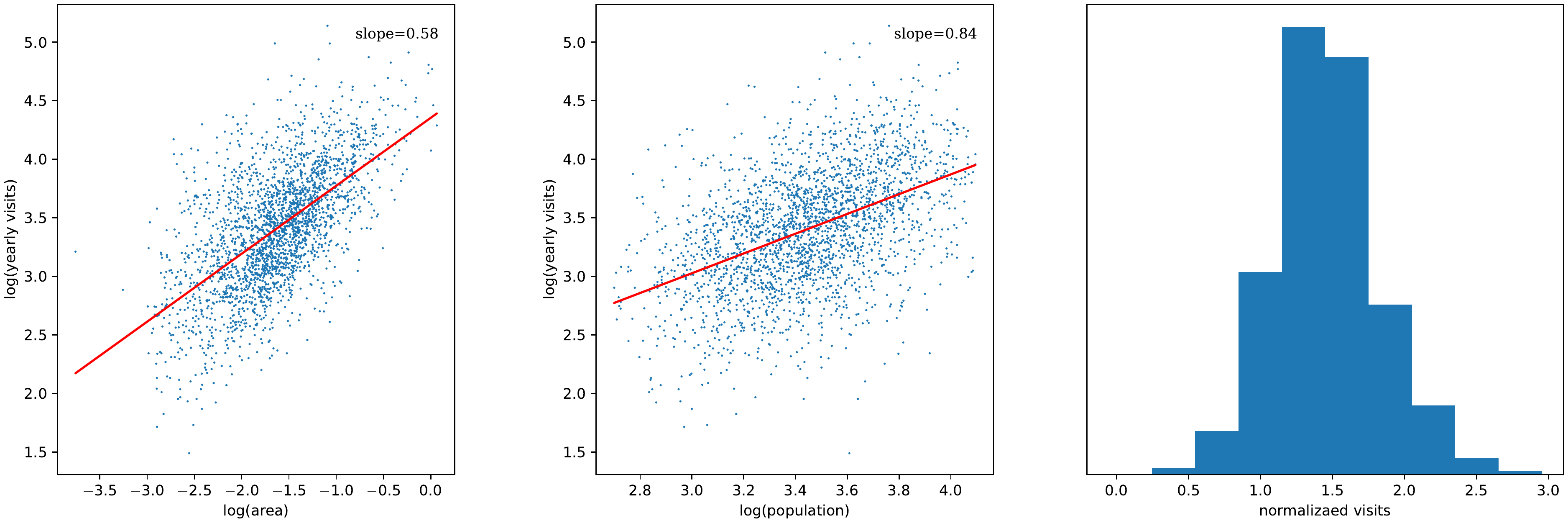}
    \caption{\label{fig:norm_fig} \textbf{Log-log plots displaying the relationship between yearly visits and park area and walkshed population, and the distribution of normalized visitation.} In log space park area is positively correlated with yearly visits, as is population of the walkshed. The slope of the line of best fit is 0.58 for park area, and 0.84 for walkshed population. These slopes were used to determine the normalized visitation value, given by Equation \ref{eq:norm}. Visitation remains log-normally distributed after normalization. See Figure \ref{fig:cartoon} for distribution of visits prior to normalization.}
\end{figure*}

To adjust for the area of the park and the walkshed population, we normalize for the increased visitation that is likely only related to the increased population to which the park is available. 
To normalize we fit simple linear regressions for park area and walkshed population in the log-log space.
The resulting slopes were used as exponents in the following normalization parameter: 

\begin{equation} \label{eq:norm}
    n_p= \frac{1}{A_p^{0.58}*P_w^{0.84}} 
\end{equation}
    
where $A_p$ is the area of the park in $\mathrm{km}^2$, and $P_w$ is the population of the walkshed. The exponents for population and area were determined using the slope of the line of best fit for each variable when plotted against yearly visits in log-log space (See Figure \ref{fig:norm_fig}).

Normalized visitation $n_p$ was used as the target value for a set of predictive models that took as training data demographic features associated with walksheds. Both input and target values were standardized by subtracting the mean and dividing by the sample standard deviation, prior to fitting, to account for different scales of measurement for the variables. 
The models tested were: Support Vector Machine Regression with linear, polynomial, and radial basis function kernels; Random Forest Regression; LASSO Regression; and Elastic Net Regression. These models were selected for the variety of functional forms offered, and for the ability of some to consider subsets of features. Random Forests were included to account for possible interaction effects between demographic features. Model fitting and analysis were performed in Python using Scikit-learn\cite{scikit-learn}.

Model hyper parameters were optimized using a grid search of the parameter space (see Table~\ref{tab:results}) over which models were evaluated using 5-fold cross validation.
The best performing parameters and the model scores are presented in Table~\ref{tab:results}. 

Model performances were compared against a null model: a constant function set to the mean of the normalized visitation. 
The null model asserts that all parks receive the same visitation 
The empirical distribution of normalized visitation suggests that parks do
not receive an equal number of annual visitors (\ref{fig:norm_fig}) and so we expect the null model to perform
poorly.

\section{\label{sec:Results} Model Performance}

\def\arraystretch{1.75}
\begin{table*}[h!]
\centering
\begin{tabular}{llllll}
\hline
\textbf{Model}~                  & \textbf{Hyperparameters }                     & \textbf{Hyperparameter Grid}                                                                        & \textbf{Tuned Hyperparameters}~ & \textbf{Mean Absolute Error} &   \\

\hline
Constant (null)~        & -                                   & -                                                                                          & -                     & 0.2810               &  \\

SVR - linear kernel     & C                                    & {[}$10^{-5},10^{-4},10^{-3},10^{-2}$]                                                              & $10^{-4}  $            & 0.2806               &   \\

SVR - rbf kernel        & \begin{tabular}[t]{@{}l@{}}C\\ $\gamma$  \end{tabular}        & \begin{tabular}[t]{@{}l@{}}{[}0.01,0.02,0.03...1] ,\\{[}$\frac{1}{n_{features}}$, $\frac{1}{(n_{features} * X.var()}$]\end{tabular}   & \begin{tabular}[t]{@{}l@{}}0.99\\ $\frac{1}{n_{features}}$  \end{tabular}           & 0.2793               &   \\

SVR - polynomial kernel &   \begin{tabular}[t]{@{}l@{}l@{}}C\\ $\gamma$ \\ degree \end{tabular}                                   &

\begin{tabular}[t]{@{}l@{}l@{}}{[}$10^{-5},10^{-4},10^{-3},10^{-2}$] ,\\{[}$\frac{1}{n_{features}}$, $\frac{1}{(n_{features} * X.var()}$] \\ {[}1,2,3,4,5]\end{tabular}   &

\begin{tabular}[t]{@{}l@{}l@{}} $10^{-1}$\\ $\frac{1}{n_{features}}$ \\ 3  \end{tabular}                                                                      & 0.2736               &   \\

LASSO                   & $\alpha$            & {[}$10^{-5},10^{-4},10^{-3},10^{-2}$]                                                             & $10^{-4} $             & 0.2759               &   \\

Elastic Net             & \begin{tabular}[t]{@{}l@{}}$\alpha$ \\ l1\_ratio \end{tabular} & \begin{tabular}[t]{@{}l@{}}{[}$10^{-5},10^{-4},10^{-3},10^{-2}$],\\{[}0.00, 1, 0.01]\end{tabular} & \begin{tabular}[t]{@{}l@{}}$10^{-4}$\\ 0.01  \end{tabular}        & 0.2758               &   \\

Random Forest           &      \begin{tabular}[t]{@{}l@{}l@{}l@{}l@{}l@{}}$n_{estimators}$\\ $max_{features}$ \\ $max_{depth}$\\ min samples to split\\ min samples at leaf\\ bootstrap \end{tabular}                               &   \begin{tabular}[t]{@{}l@{}l@{}l@{}l@{}l@{}} {[}200, 400]\\ $n_{features}, \sqrt{n_{features}}$ \\ {[}10,30,50,70,90,110]\\ {[}2,5,10]\\ {[}1,2,4]\\ {[}True, False] \end{tabular}                                                                                          &  \begin{tabular}[t]{@{}l@{}l@{}l@{}l@{}l@{}} 200\\ $\sqrt{n_{features}}$ \\10\\2\\ 1\\ False \end{tabular}                     &  0.2664                  &   \\

\hline
\end{tabular}

\caption{  \label{tab:results} \textbf{Results of training and testing predictive models for predicting normalized yearly park visitation using the demographic features associated with that park's walkshed.} Model types are presented along with the hyperparameters they use. The hyperparameters were tuned using a 5 fold cross-validation and a grid search over the parameter space. For each model the grid used for the grid search is reported followed by the tuned hyperparameters and the Mean Absolute Error of the tuned model.  The notation $n_{features}$ refers to the number of independent variables, and $X.var()$ refers to the variance of the feature array. The l1\_ratio is the ratio of the weights of the $L1$ and $L2$ penalties, such that l1\_ratio $=1$ is the LASSO penalty. All hyperparameters refer to those in the scikit-learn library \cite{scikit-learn}.
}
\end{table*}

Modeling the data using a constant function set to the mean of the normalized visitation yielded a Mean Absolute Error (MAE) of 0.2810. The random forest model was most successful of the regression models, achieving a MAE of 0.2664, or an 5.20\% improvement over the null model.

Support Vector Regression (SVR) with a polynomial kernel was the second most accurate predictive model. This model resulted in a MAE of 0.2746, which is 2.33\% less than the null model. SVR  using radial basis function kernels had slightly worse performance, obtaining a MAE of 0.2793; only a 0.60\% improvement over the null model. When paired with a linear kernel SVR performed comparably to the null model, with a MAE of 0.2810.

The other two linear models performed equally poorly.
LASSO had a minimum MAE of 0.2759, and Elastic Net managed a very similar minimum MAE of 0.2758, indicating that they probably converged to very similar linear models.
This is only an 1.81\%, and 1.85\% improvement upon the null model. 

Given that the most successful model, the random forest, only achieved an improvement of  5.20\% over the null model, we can conclude that demographic information did not significantly contribute to more accurate predictions of normalized yearly visits.

\section{\label{sec:discussion}Discussion}

Our approach investigated a variety of functional forms, employed methods for decreasing noise created by unimportant features (LASSO and Elastic Net), and incorporated the potential for interaction effects through sequential variable consideration in the Random Forest model.
Additionally, the hyperparameters of each model were tuned. None of the predictive models performed substantially better than the null model. The null model provides no information about individual park usage, and has very weak predictive power. It also assumes that the demographic features considered in this study are irrelevant to visitation prediction, as it does not use them at all. Since a comprehensive set of models was tested, and none performed considerably better than the null model, we can infer that the demographic features have little predictive value for average yearly park visitation. 

This result is perhaps unexpected in light of Hamstead et al's 2018 findings in New York City \cite{hamstead_geolocated_2018}. In contextualizing this discrepancy, it is important to note that our inclusion criteria  focuses on community parks in suburban or urban residential areas, while previous work focused on parks in general. This effort decreased the influence of tourism on visitation by decreasing the number of parks that would be considered tourist destinations. In addition, exclusion of these parks naturally excludes potential demographic disparity in residential proximity to destination parks (i.e. homes near destination parks may be more expensive) as a confounding effect. Thus it is possible that our results differ because of the type of parks considered.

This study is limited by the inability to determine the residential proximity of visitors. We observed a correlation between the number of people in a park's walkshed and the number of visits received by a park. This suggests that the number of people a park ``serves" is related to the number of people who visit it. In order to control for the effect of park size, and geographical accessibility, a normalization constant was applied to control for this relationship. In doing so, we made an assumption that a ten minute walk to the park encompassed the `service area' of every park. It is possible that depending on demographic features, some populations may travel farther to parks on average than others, but maintain similar visitation rates. If this were the case, the normalization constant for parks serving those populations would be too small, inflating the normalized visitation observed, and potentially obscuring important trends. 

It is worth reiterating that these results are park-centered; results speak to the usage that a park receives, but do not reflect \textit{who} uses the park. Our data does not provide the origin of the mobile device visiting the park there is no way to determine if visits are made by the people living in the walkshed, or by persons living farther away. Therefore, whether park usage differs for different \textit{populations} is not addressed by this work.

Future work should include more detailed data on the origin of the devices observed in each park. This data would allow for a more accurately constructed walkshed. It would also allow further exploration into who visits parks, and which parks they visit. 
In addition, consideration should be given to park quality; this work did not incorporate features of the park itself into the predictive modeling. Park amenities, condition, and environment are all important contributors to visitor attraction. Since aspects of park quality could be confounding factors with walkshed demographic features, a logical extension of this work is to control for this confounding.

Further future analyses would benefit from consideration of park visitation relative to season and weather. In the current study yearly visitation is used, which obscures the difference in ``visitation-season" length between parks; parks in Southern California may be used more days in a year relative a park in Boston, which creates a different story of population and visitation.

\raggedright


%

\end{document}